\documentclass[a4paper, 11pt]{article}
\usepackage{graphicx, amssymb}
\newif\iffigs
\figstrue   
\iffigs
\fi
\def\drawing #1 #2 #3 {
\begin{center}
\setlength{\unitlength}{1mm}
\begin{picture}(#1,#2)(0,0)
\put(0,0){\framebox(#1,#2){#3}}
\end{picture}
\end{center} }
\newcommand{\eqref}[1]{(\ref{#1})}

\newcommand{\figref}[1]{Fig.~\ref{#1}}

\newcommand{\Prob}[0]{{\rm Prob}}
\newcommand{\pstar}[0]{{p_\star}}

\newcommand{\D}[2]{\Delta^{({#1})}_{{#2}}}
\newcommand{\Dle}[2]{\Delta^{({#1})}({#2})}
\newcommand{\nset}{\mathbb{N}}
\newcommand{\rset}{\mathbb{R}}

\def\be{\begin{equation}}
\def\ee{\end{equation}}
\def\p{\prime}

\begin{document}
\title{Multifractality of the Feigenbaum attractor and fractional 
derivatives}
\author{U. Frisch$^{1}$, K. Khanin$^{1-4}$ and T. Matsumoto$^{1,5}$ \\
$^1${\small Observatoire de la C\^{o}te d'Azur,  B.P. 4229, 06304 Nice Cedex 4,
France}\\
$^2${\small Issac Newton Institute for Mathematical Sciences, 20 Clarkson Road,
Cambridge CB3 0EH, U.K.}\\
$^3${\small Dept. of Mathematics, Heriot-Watt University, Edinburgh EH14
4AS, U.K.}\\
$^4${\small Landau Institute for Theoretical Physics, Kosygina Str. 2, Moscow
117332, Russia.}\\
$^5${\small Dept. of Physics, Kyoto University, Kitashirakawa Oiwakecho
Sakyo-ku Kyoto 6068502, Japan.}}
\date{\today}

\maketitle

\centerline{\it J.\ Stat.\ Phys. in press}
\begin{abstract}
 It is shown that fractional derivatives of the (integrated) invariant
measure of the Feigenbaum map at the onset of chaos have
power-law tails in their cumulative distributions, whose exponents can
be related to the spectrum of singularities $f(\alpha)$.  This is a
new way of characterizing multifractality in dynamical systems, so far
applied only to multifractal random functions (Frisch and Matsumoto
(\textit{J.\ Stat.\ Phys.} \textbf{108}:1181, 2002)).  The relation
between the thermodynamic approach (Vul, Sinai and Khanin
(\textit{Russian Math.\ Surveys} \textbf{39}:1, 1984)) and that based
on singularities of the invariant measures is also examined.  The
theory for fractional derivatives is developed from a heuristic point
view and tested by very accurate simulations.
\end{abstract}
\begin{center}
{\bf Keywords:}
chaotic dynamics, multifractals, thermodynamic formalism.
\end{center}

\section{Introduction}
\label{s:intro}

Recently a new method for analyzing multifractal functions $u(x)$ 
was introduced  \cite{fm02}. It exploits 
the fact that the fractional derivative of order $a$ (denoted here by $D^{a}$)
of  $u(x)$ has, for a suitable range of $a$,
a power-law tail in its cumulative probability 
\begin{equation}
 \Prob\{|D^{a}u| > \xi \} \propto \xi^{-\pstar}, \quad \xi \to \infty.
\label{e:power-law-tail}
\end{equation}
The exponent $\pstar(a)$ is the unique solution of the equation
\begin{equation}
    \zeta_{p_\star} =\pstar(a) a,
    \label{e:pstar}
\end{equation}
where $\zeta_p$ is the scaling exponent associated to the behavior at small
separations $l$ of the structure function of order $p$, i.e. $\langle |u(x +
l) - u(x)|^p\rangle \sim |l|^{\zeta_p}$. It was also shown that the actual
observability of the power-law tail when multifractality is restricted to a
finite range of scales is controlled by how much $\zeta_p$ departs from 
linear dependence  on $p$. The larger this departure 
the easier it is to observe multifractality.

So far the theory of such power-law tails has been developed
only for synthetic random functions, in particular the
random multiplicative process \cite{benzi93} for which Kesten-type maps
\cite{kesten73} and large deviations theory can be used.

It is our purpose here to test the fractional derivative method for
invariant measures of dissipative dynamical systems, in particular for
the Feigenbaum invariant measure which appears at
the accumulation point of the period doubling cascade where the orbit
has period $2^{\infty}$ \cite{f78, f80}.
Its multifractality was proven rigorously in 
Ref.~\cite{vsk84} using a thermodynamic formalism. For the 
Feigenbaum measure  
all scaling exponents can be determined with arbitrary accuracy. 

There is an important difference in the way one processes 
functions and invariant measures to determine their multifractal 
properties and in particular the spectrum  of
singularities, usually denoted $D(h)$ for functions \cite{pf85}
and $f(\alpha)$ for measures \cite{bppv84, HJKPS}. For a function 
$u(x)$ one uses the moments or the PDFs of the increments $u(x+l) - u(x)$ 
to determine the scaling exponents, whereas for an invariant  measure
$\mu_0$ one works with integrals over intervals or boxes of different 
sizes. In the one-dimensional case the two approaches become equivalent 
by introducing the cumulative distribution function 
\begin{equation}
\label{distr}
u(x) \equiv \int_{-\infty}^{x} \mu_0(dx).
\end{equation} 
Hence we shall apply the fractional derivative method to the integral
of the invariant measure.

The organization of the paper is the following. Section~\ref{s:thermo}
is devoted  to  the thermodynamic
formalism for the Feigenbaum attractor.
In Section~\ref{ss:formalism}, we recall the method used in Ref.~\cite{vsk84}.
In Section~\ref{ss:connection} we show how this formalism, 
based on the study of the geometrical properties 
of the attractor, is
actually connected to the standard multifractal 
formalism which focusses on the statistical properties of the invariant measure
\cite{bppv84, HJKPS}. To the best of our 
knowledge the exact relation between the two formalisms is discussed here
for the first time. 
Then, in Section~\ref{ss:numericalfreeenergy}
we calculate numerically the free energy and accordingly
the scaling exponents $\zeta_p$ for the integral of the invariant measure;
this is done by a very accurate transfer-matrix-based method.
Fractional derivatives are discussed in Section~\ref{s:fraclap}.
In Section~\ref{ss:fraclap_pheno} we briefly recall the 
phenomenology of power-law tails in the distribution of fractional 
derivatives and the limits on observability. The fractional derivative 
analysis of the Feigenbaum measure is presented
in Section \ref{ss:fraclap_numerics}.
Concluding remarks are made in Section~\ref{s:concl}.

\section{Thermodynamic Formalism for the Feigenbaum Attractor}
\label{s:thermo}
\subsection{Thermodynamic formalism}
\label{ss:formalism}
In this section we give a brief description of the thermodynamic formalism for
the invariant measure of the Feigenbaum map (see Ref.~\cite{vsk84} for
the mathematical details) and show how one can use it in order to study the
multifractal properties of the H\"older exponents.

By Feigenbaum attractor we understand the attractor of the one-dimensional
mapping $g : [0,1] \to [0,1]$ , where $g(x)$ is the solution of
the Feigenbaum--Cvitanovi\'{c} doubling equation:
\begin{equation}
g(x) = -\frac{1}{\gamma}g(g(\gamma x)) \, , \,\,\, g(0)=1 \, , \,\,\, \gamma = \frac{1}{\alpha} = - g(1) \, .
\label{g}
\end{equation}
Equation (\ref {g}) is known to have the unique solution in the class of smooth
unimodal maps (that is, maps having one critical point) with a non-degenerate maximum.
This solution is called the Feigenbaum map. It is holomorphic in some
complex neighborhood of $[-1,1]$ and the first few terms in the power
series expansion are \cite{l82}
\begin{equation}
g(x)=1 - 1.5276\dots x^2 + 0.1048\dots x^4 + 0.0267\dots x^6 -
 0.003527\dots x^8 + \dots \, .
\label{g1}
\end{equation}
The value of the universal constant $\gamma$ which is the inverse of the 
Feigenbaum 
scaling constant $\alpha$ is approximately equal to $0.3995$.
An attractor $A$ for the map $g$ can be constructed in the following way.
For each $n\geq 1$ define a collection of intervals of $n-$th level:
\begin{eqnarray}
&&\Delta^{(n)}_0 = [-\gamma^n,\gamma^n], \nonumber \\
&&\Delta^{(n)}_i = g^{(i)}(\Delta^{(n)}_0) \equiv \underbrace{g \circ g
 \circ \cdots \circ g}_{i} (\Delta_0^{(n)}) \quad (1 \leq i\leq 2^n-1).
\label{delta}
\end{eqnarray}

\begin{figure}
\iffigs 
\centerline{%
\includegraphics[scale=0.7]{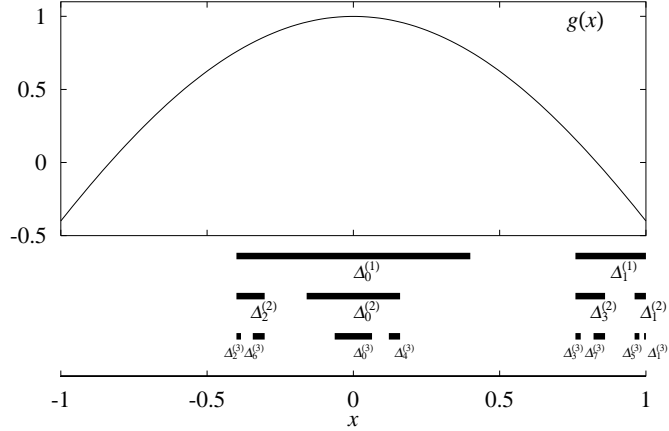}
}
\else\drawing 65 10 {dynamical partitions}
\fi
\caption{\label{f:dynamicalpartition} The Feigenbaum--Cvitanovi\'c map $g(x)$
 and the first three levels of the partitions $\D{n}{j}$. For $g(x)$ we used
 the expansion \eqref{g1}, introduced in Ref.~\cite{l82} up to $x^{32}$.}
\end{figure}

The following properties of the intervals $\Delta^{(n)}_i$ are easy
consequences of the doubling equation (\ref {g}):
(a) Intervals $\Delta^{(n)}_i, 0 \leq i\leq 2^n-1$ are pairwise disjoint.
(b) $g^{(2^n)}\Delta^{(n)}_0 \subset \Delta^{(n)}_0$.
(c) Each interval of $n-$th level $\Delta^{(n)}_i$ contains exactly two intervals of $(n+1)-$th level,
$\Delta^{(n+1)}_i$ and $\Delta^{(n+1)}_{i+2^n}$.
(d) $2\gamma^{2n}\leq |\Delta^{(n)}_i|\leq 2\gamma^n$, where $|\cdot|$
denotes the length of the interval.
The first three levels of the intervals are shown in Fig.~\ref{f:dynamicalpartition}.

The properties above imply that it is natural to use a dyadic representation
for the intervals $\Delta^{(n)}_i$. Let $i = \sum_{j=0}^{n-1}\epsilon_j \,
2^i$, where $\epsilon_j = 0,1$.  Then we can use a sequence $(\epsilon_0,
\dots, \epsilon_{n-1})$ as a symbolic coding for intervals $\Delta^{(n)}_i$:
$\Delta^{(n)}_i = \Delta^{(n)}_{\epsilon_0, \dots, \epsilon_{n-1}}$.  Now we
can define the Feigenbaum attractor
\begin{equation}
A=\bigcap_{n\geq 1} \bigcup_{i = 0}^{2^n-1} \Delta^{(n)}_i \, .
\label{F}
\end{equation}
The set $A$ is isomorphic to the set of all infinite dyadic sequences $(\epsilon_0,\epsilon_1, \dots, \epsilon_{n}, \dots)$.
Such sequences can be considered as a symbolic coordinate system on $A$.
In this new coordinate system the map $g$ acts as the dyadic addition of the sequence $(1,0, \dots, 0, \dots)$.
Notice that topologically $A$ is a Cantor set. It is easy to see that $A$ is indeed
an attractor for all but countably many initial points $x \in [-1,1]$:
${\rm dist}(g^{(m)}(x) , A) \to 0$ as $m \to \infty$. The exceptional set of initial points
consists of all unstable periodic orbits and their preimages.

As we have seen above, all intervals $\Delta^{(n)}_i$ have exponentially small
lengths but the exponent varies from $2\gamma^n$ to $\gamma^{2n}$. Notice that
exponents $\ln |\Delta^{(n)}_i|$ give all possible scalings of the fractal set $A$.
The basic ingredient
which is needed for the multifractal analysis is the control over the spectrum
of possible scalings corresponding to exponents, i.e. $\ln |\Delta^{(n)}_i|$. Such control can be achieved with the help
of the thermodynamic formalism. The thermodynamic formalism which was constructed in Ref.~\cite {vsk84} is based
on the Gibbsian description for the lengths of the intervals $\Delta^{(n)}_i$. It is shown in
Ref.~\cite {vsk84} that there exists a function $U(\epsilon^{(1)}, \epsilon^{(2)}, \dots, \epsilon^{(n)}, \dots)$ (thermodynamic potential)
which is defined on all infinite dyadic sequences such that:

I. There exists a constant $C >0$ for which
\begin{equation}
\label{c}
\frac{1}{C} \leq -U(\epsilon^{(1)}, \epsilon^{(2)}, \dots, \epsilon^{(n)}, \dots)\leq C \, .
\end{equation}

II. For any two dyadic sequences
$$(\epsilon^{(1)}, \epsilon^{(2)}, \dots, \epsilon^{(n)}, \bar{\epsilon}^{\,(n+1)}, \bar{\epsilon}^{\,(n+2)}, \dots) \, , \, (\epsilon^{(1)}, \epsilon^{(2)}, \dots, \epsilon^{(n)}, \tilde{\epsilon}^{\,(n+1)}, \tilde{\epsilon}^{\,(n+2)}, \dots)$$

\ \ \ \ which coincide on the first $n$ positions
\begin{equation}
\label{approx}
|U(\epsilon^{(1)}, \epsilon^{(2)}, \dots, \epsilon^{(n)}, \bar{\epsilon}^{(n+1)}, \bar{\epsilon}^{\,(n+2)}, \dots) - U(\epsilon^{(1)}, \epsilon^{(2)}, \dots, \epsilon^{(n)}, \tilde{\epsilon}^{\,(n+1)}, \tilde{\epsilon}^{\,(n+2)}, \dots)|\leq C(2\gamma)^n \, .
\end{equation}

III. For any $\Delta^{(n)}_{\epsilon_0, \dots, \epsilon_{n-1}}$ with $\epsilon_0 = 1$
\begin{equation}
\label{exp1}
\exp[-C(2\gamma)^{\frac{n}{3}}]\leq \frac{|\Delta^{(n)}_{\epsilon_0, \dots, \epsilon_{n-1}}|}{|\Delta^{(n-1)}_{\epsilon_0, \dots, \epsilon_{n-2}}|}\exp(-U(\epsilon_{n-1}, \dots, \epsilon_{1},1,0,\dots ,0,\dots ))\leq \exp[C(2\gamma)^{\frac{n}{3}}] \, .
\end{equation}
It immediately follows from (\ref{exp1}) that for $C_1 = \exp \left[\frac{C}{1-(2\gamma)^{\frac{1}{3}}}\right]$
\begin{equation}
\label{exp2}
\frac{1}{C_1}\leq
\frac{|\Delta^{(n)}_{1,\epsilon_1, \dots, \epsilon_{n-1}}|}
{\exp\left[\sum_{s=0}^{n-1}U(\epsilon_s, \epsilon_{s-1}, \dots, \epsilon_1,1,0,\dots, 0, \dots)\right]}\leq C_1 \, .
\end{equation}
The condition $\epsilon_0=1$ corresponding to intervals $\Delta_i^{(n)}$ with
odd $i$'s plays only a technical role and it is not essential for our further
analysis since the odd intervals contain information about the lengths of the
even ones. Indeed, it is very easy to see that for every odd $i$ the intervals
$\Delta_i^{(n)}$ and $\Delta_{i+1}^{(n)}$ have lengths of the same order.

We next introduce a parameter $\beta$ (inverse temperature) and define the
partition function
\begin{equation}
Z_n(\beta) = \sum_{\epsilon_{n-1}, \dots, \epsilon_0 = 0,1}\exp\left[\beta \sum_{s=0}^{n-1}U(\epsilon_s, \epsilon_{s-1}, \dots, \epsilon_1,1,0,\dots, 0, \dots)\right]
\label{part}
\end{equation}
and the free energy
\begin{equation}
F_n(\beta) = \frac{\ln Z_n(\beta)}{n} \, , \,\,\,\,\, F(\beta) = \lim_{n \to \infty}F_n(\beta) \, .
\label{free}
\end{equation}
It immediately follows from (\ref{part}) and (\ref{free}) that
\begin{equation}
\label{free1}
\sum_{i=0}^{2^n-1}|\Delta_i^{(n)}|^\beta \, \sim \,  \exp[nF(\beta)].
\end{equation}

In the thermodynamic limit $n \to \infty$ the probability distributions
\begin{equation}
\nu^{(n)}_\beta (\epsilon_{n-1}, \epsilon_{n-2}, \dots, \epsilon_1)=\frac{1}{Z_n(\beta)}\exp\left[\beta\sum_{s=0}^{n-1}U(\epsilon_s, \epsilon_{s-1}, \dots, \epsilon_1,1,0,\dots, 0, \dots)\right]
\label{Gibbs}
\end{equation}
tend to a limiting distribution $\nu_\beta$ which can be considered as a Gibbs
measure with the potential $U$, inverse temperature $\beta$ and
the boundary condition $\epsilon_{0} = 1, \,\, \epsilon_{-i} = 0, \,\, i
\in \nset$.
This Gibbs distribution generates the probability measure on
$A_+ = A \cap \Delta^{(1)}_1$ which is the part of the whole attractor
$A$ corresponding to intervals $\Delta^{(n)}_i$ with odd numbers $i$.
We shall denote this Gibbs measure on $A_+$ by $\mu_{\beta}$.
Notice that $\beta=0$ corresponds to a unique invariant measure and $\beta =1$ gives a
conditional distribution corresponding to Lebesgue measure on $[-1,1]$.  The
free energy $F(\beta)$ contains all information about the multifractal
properties of the Feigenbaum attractor.
Notice that the thermodynamic formalism leads to
one-dimensional statistical mechanics with exponential decay of
interactions and hence without phase transitions.
This implies that $F(\beta)$ is a smooth function. In fact it is holomorphic
in some complex neighborhood of the real axis. Denote
\begin{equation}
H(\epsilon_{n-1}, \dots, \epsilon_1) = \sum_{s=0}^{n-1}U(\epsilon_s, \epsilon_{s-1}, \dots, \epsilon_1,1,0,\dots) \, .
\end{equation}
Using relations
\begin{eqnarray}
\label{monotonicity}
F^{\p}_n(\beta)&=&\frac{1}{Z_n(\beta)}\sum_{\epsilon_{n-1}, \dots, \epsilon_0 = 0,1}\frac{1}{n}H(\epsilon_{n-1}, \dots, \epsilon_1)
\exp\left[\beta H(\epsilon_{n-1}, \dots, \epsilon_1)\right] \nonumber \\
&=&\left<\frac{\ln |\Delta^{(n)}_{1,\epsilon_1, \dots, \epsilon_{n-1}}|}{n}\right>_{\nu^{(n)}_\beta} + O\left(\frac{1}{n}\right)
\end{eqnarray}
and
\begin{eqnarray}
\label{convexity}
F^{\p \p}_n(\beta)&&=\frac{1}{Z_n(\beta)}\sum_{\epsilon_{n-1}, \dots, \epsilon_0 = 0,1}\left(\frac{1}{n}H(\epsilon_{n-1}, \dots, \epsilon_1)\right)^2
\exp\left[\beta H(\epsilon_{n-1}, \dots, \epsilon_1)\right] \nonumber \\
&&-\left(\frac{1}{Z_n(\beta)}\sum_{\epsilon_{n-1}, \dots, \epsilon_0 = 0,1}\frac{1}{n}H(\epsilon_{n-1}, \dots, \epsilon_1)
\exp\left[\beta H(\epsilon_{n-1}, \dots, \epsilon_1)\right]\right]^2 \nonumber \\
&&=\left<\left(\frac{\ln |\Delta^{(n)}_{1,\epsilon_1, \dots,
	  \epsilon_{n-1}}|}{n}\right)^2\right>_{\nu^{(n)}_\beta} -
\left(\left<\frac{\ln |\Delta^{(n)}_{1,\epsilon_1, \dots,
       \epsilon_{n-1}}|}{n}\right>_{\nu^{(n)}_\beta}\right)^2 
       + O\left (\frac{1}{n}\right),
\end{eqnarray}
we conclude that $F(\beta)$ is a monotone decreasing convex function.
We shall also use the spectral representation for
the free energy. Consider the transfer-matrix operator ${\cal L}(\beta)$:
\begin{equation}
\label{transfer}
{\cal L}(\beta)h(\epsilon^{(1)}, \epsilon^{(2)}, \dots, \epsilon^{(n)}, \dots) = \sum_{\epsilon^{(0)} = 0,1}\exp[\beta U(\epsilon^{(0)},\epsilon^{(1)}, \epsilon^{(2)}, \dots, \epsilon^{(n)}, \dots)]h(\epsilon^{(0)},\epsilon^{(1)}, \epsilon^{(2)}, \dots, \epsilon^{(n)}, \dots) \, .
\end{equation}
Since ${\cal L}(\beta)$ is a positive linear operator,its largest eigenvalue $\lambda(\beta)$ is strictly positive and simple.
It is easy to see that
\begin{equation}
\label{specfree}
F(\beta) = \ln \lambda(\beta)\, .
\end{equation}

For an arbitrary
point $x \in A_+$, denote by $\Delta^{(n)}(x)$ the interval of the $n$th level
which contains $x$. It follows from (\ref {monotonicity}) that for points
$x$ which are typical with respect to $\mu_{\beta}$ (that is
corresponding to a set of full $\mu_{\beta}$-measure.)
\begin{equation}
\label{typ1}
|\Delta^{(n)}(x)| \, \sim \,   \exp[F^{\p}(\beta)n] \, .
\end{equation}
More precisely, for $\mu_{\beta}$-almost all $x \in A$
\begin{equation}
\label{typ2}
\lim_{n \to \infty} \frac{\ln|\Delta^{(n)}(x)|}{n} = F^{\p}(\beta ) \, .
\end{equation}
We next find the total number $N_n(\beta)$ of the intervals of $n-$th level
whose length is of the order $|\Delta^{(n)}(\beta)|=\exp[F^{\p}(\beta)n]$.
We have
\begin{equation}
N_n(\beta)|\Delta^{(n)}(\beta)|^\beta \,
 \sim
\,  N_n(\beta)\exp[\beta F^{\p}(\beta)n]\,
\sim
\,  Z_n(\beta) \, \sim \,  \exp[F(\beta)n]
\end{equation}
which gives
\begin{equation}
\label{N}
N_n(\beta)\, \sim \,  \exp[(F(\beta)-\beta F^{\p}(\beta))n] \, .
\end{equation}
Using (\ref {N}) we can find the Hausdorff dimension $d_H(\beta)$ of the set of
points $x \in A_+$ which are typical with respect to the measure $\mu_\beta$.
Since
\begin{equation}
\label{dim1}
N_n(\beta)|\Delta^{(n)}(\beta)|^{d_H(\beta)} \, \sim \,  \exp[(F(\beta)-\beta F^{\p}(\beta) +d_H(\beta)F^{\p}(\beta))n]
\end{equation}
we conclude that
\begin{equation}
F(\beta)-\beta F^{\p}(\beta) +d_H(\beta)F^{\p}(\beta)=0
\end{equation}
which immediately implies
\begin{equation}
\label{dim2}
d_H(\beta)=\frac{\beta F^{\p}(\beta)-F(\beta)}{F^{\p}(\beta)}= \beta - \frac{F(\beta)}{F^{\p}(\beta)} \, .
\end{equation}
The Hausdorff dimension $d_H(A)$ of the whole attractor $A$ is equal to the
maximum of $d_H(\beta)$ over all $\beta \in \rset$. Let $\beta_0$ be the
unique solution of the equation $F(\beta)=0$. It is easy to see that $d_H(A) =
d_H(\beta_0)=\beta_0$.

\begin{figure}
\iffigs 
\centerline{%
\includegraphics[scale=0.8]{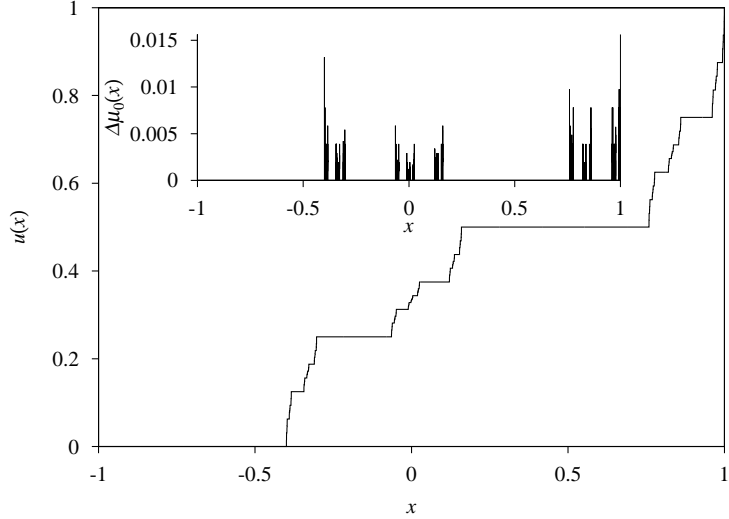}
}
\else\drawing 65 10 {feigenbaum measure, u(x)}
\fi
\caption{\label{f:feigendistribution}
 The integral $u(x)$ of the Feigenbaum invariant measure calculated with 
$2^{15}$ bins of  uniform length $\Delta = 2^{-14}$ in $[-1,1]$. 
Inset: The invariant measure smoothed over the distance $\Delta$ 
calculated as a frequency histogram.
}
\end{figure}
We next discuss multifractal properties associated with the H\"older exponents.
Consider the integral $u(x)$ of the invariant measure 
$\mu_0$,
defined by \eqref{distr}, 
which is plotted in \figref{f:feigendistribution}. The attractor being
topologically a Cantor set, 
 $u(x)$ is a variant of the Devil's staircase (see Ref.~\cite{frisch95},
 Section 8.2).

To find its spectrum of H\"older exponents, notice that for every interval $\Delta^{(n)}_i$ the increase of $u(x)$ along the interval
is equal to $2^{-n}$. Hence $\Delta^{(n)}_i$
corresponds to a H\"older exponent $h_n(i)=\alpha$ where
$|\Delta^{(n)}_i|^{\alpha} \, \sim \,  \frac{1}{2^n}$.
This implies
\begin{equation}
|\Delta^{(n)}_i| \, \sim \,  2^{-\frac{n}{\alpha}} \, .
\label{hold}
\end{equation}
Using (\ref {typ1}), (\ref {hold}) we conclude that the H\"older exponent
$\alpha$ corresponds to an inverse temperature $\beta(\alpha)$ such that
\begin{equation}
F^{\p}(\beta(\alpha))= - \frac{\ln 2}{\alpha} \, .
\end{equation}
This gives 
\begin{equation}
\beta(\alpha) = (F^{\p})^{-1}\left(-\frac{\ln 2}{\alpha}\right) \, ,
\label{betah}
\end{equation}
where $(F^{\p})^{-1}(y)$ is the inverse function to $F^{\p}(x)$.
We can now find the Hausdorff dimension $d_H(\alpha)$ of the set of points $x \in A$
for which the H\"older exponent of $u(x)$ is equal to $\alpha$:
\begin{equation}
\label{dh}
d_H(\alpha) = d_H(\beta(\alpha))=\beta(\alpha) - \frac{F(\beta(\alpha))}{F^{\p}(\beta(\alpha))} \, .
\end{equation}
Notice that the analysis presented above can be made completely rigorous
(see, for example, Refs.~\cite{vsk84, cla87}).

\subsection{Connection between  thermodynamic formalism and 
standard multifractal analysis}
\label{ss:connection}

It is quite interesting to compare the multifractal analysis which we
presented above with the one introduced in Ref.~\cite{HJKPS}.
Although we restrict ourselves here to the case of the Feigenbaum
attractor, the result presented below  holds in a much more general
setting. Basically our analysis is valid whenever
the system under consideration can be described with the help of the
thermodynamic formalism.

The basic object for our analysis is the Feigenbaum attractor itself and
the method is based on
the construction of the thermodynamic formalism for the lengths of the
elements of dynamic partitions $\Delta^{(n)}_i$. The thermodynamic formalism
uses considerable dynamical information about the map $g$. In contrast, the
analysis in Ref.~\cite{HJKPS} is carried out for fractal measures and does not
directly use the dynamical information about the system. In the
period-doubling setting the fractal measure is $\mu_0$. It is the unique
invariant measure for $g$ acting on $A$ (see Ref.~\cite{vsk84}).
It also can be considered as a
physical or Sinai--Ruelle--Bowen (SRB) measure on $[-1,1]$. This means that
under dynamics given by the map $g$ any initial absolutely continuous
distribution $\bar \mu$ on $[-1,1]$ converges to $\mu_0$: $g^{(n)} ({\bar
\mu}) \to \mu_0$ as $n \to \infty$.  The multifractal analysis in
Ref.~\cite{HJKPS} is based on a function $D_p$ which can be defined in the
following way. Consider a partition of the interval $[-1,1]$ into subintervals
$I_j,\,\, 1\leq j \leq N$ of length $l$. Then
\begin{equation}
\label{HJKPS1}
D_p=\lim_{l\to 0}\left(\frac{1}{p-1}\frac{\ln \sum_{j=1}^N
		  (\mu_0(I_j))^p}{\ln l}\right)
= \frac{\alpha}{\ln 2}F(\beta(\alpha))\, .
\end{equation}
It follows from (\ref {HJKPS1}) that
\begin{equation}
\label{HJKPS11}
\sum_{j=1}^N (\mu_0(I_j))^p \, \sim \,  N^{D_p(1-p)}\, .
\end{equation}
Another characteristic of a multifractal measure is given by its spectrum of dimensions
$f(\alpha)$ which is just the Legendre transform of $D_p(p-1)$:
\begin{equation}
\label{HJKPS2}
f(\alpha) = \inf_p \ [\alpha p - D_p(p-1)] \, .
\end{equation}
The dual Legendre relation allows one to find $D_p(p-1)$ from $f(\alpha)$:
\begin{equation}
\label{HJKPS22}
D_p(p-1) = \inf_{\alpha} \ [\alpha p - f(\alpha)] \, .
\end{equation}
We next find a correspondence between the pair $(D_p(p-1),\, f(\alpha))$ and
the pair of thermodynamic functions $(F(\beta), \,d_H(\alpha))$. We shall show
that
\begin{equation}
\label{HJKPS3}
D_p(p-1) = - F^{-1}(p\ln 2) \, , \quad f(\alpha) = d_H (\alpha) \, ,
\end{equation}
where $F^{-1}$ is an inverse function to the free energy $F(\beta)$.
To derive the first relation we consider the dynamical partition $\Delta_i^{(n)},
0\leq i \leq 2^n-1$ and assume that $n \gg 1$ but $l \ll \min_i |\Delta_i^{(n)}|$.
For each $\Delta_i^{(n)}$ define
\begin{equation}
\label{HJKPS4}
M(\Delta_i^{(n)})=\sum_{j : I_j \subset \Delta_i^{(n)}} (\mu_0(I_j))^p \, .
\end{equation}
Notice that the asymptotic behavior of $M(\Delta_i^{(n)})$ depends only on
asymptotic scalings of smaller elements of the dynamical partitions $\Delta_s^{(n+m)}$
inside $\Delta_i^{(n)}$. The thermodynamic formalism constructed above implies
that asymptotically those scalings are completely determined by the potential
$U$ and hence they do not depend on $i$. Rescaling the invariant measure inside
$\Delta_i^{(n)}$ by a factor $2^n$ we conclude that
\begin{equation}
\label{HJKPS5}
M(\Delta_i^{(n)})= \frac{1}{2^{np}}\sum_{j : I_j \subset \Delta_i^{(n)}}(2^n\mu_0(I_j))^p
\, \sim \,  \frac{1}{2^{np}}N_i^{D_q(1-p)}\, ,
\end{equation}
where $N_i$ is a total number of the intervals $I_j$ inside $\Delta_i^{(n)}$.
Taking the sum over $i$ and using (\ref{free1}) we have
\begin{eqnarray}
\label{HJKPS6}
\sum_{j=1}^N (\mu_0(I_j))^p= \sum_{0\leq i \leq 2^n-1}M(\Delta_i^{(n)}) 
&\sim &\frac{1}{2^{np}}\sum_{0\leq i \leq 2^n-1}N_i^{D_p(1-p)} \, \sim \,  \frac{1}{2^{np}}\sum_{0\leq i \leq 2^n-1}\left (\frac{|\Delta_i^{(n)}|}{l}\right)^{D_p(1-p)}\nonumber \\
&\sim& \frac{1}{2^{np}}N^{D_p(1-p)}\exp[nF(D_p(1-p))]\, .
\end{eqnarray}
This together with (\ref {HJKPS11}) immediately gives 
\begin{equation}
\label{HJKPS7}
\exp[F(D_p(1-p))]=2^p
\end{equation}
which implies the first relation in (\ref{HJKPS3}). We next show that the second relation holds.
Using (\ref{HJKPS2}) we have
\begin{eqnarray}
\label{HJKPS8}
f(\alpha) &=& \inf_p \ [\alpha p - D_p(p-1)]=\inf_p \ [\alpha p - (-F^{-1}(p\ln 2)]=\inf_z \ \left[\frac{\alpha}{\ln 2}z + F^{-1}(z)\right] \nonumber \\
&=& \inf_\beta \ \left[\frac{\alpha}{\ln 2}F(\beta) + \beta \right] = \frac{\alpha}{\ln 2}\inf_\beta\ \left[\frac{\ln 2}{\alpha}\beta + F(\beta)\right]      \, .
\end{eqnarray}
It is easy to see that the extremum in (\ref {HJKPS8}) corresponds to 
$\beta(\alpha) = (F^{\p})^{-1}\left(-\frac{\ln 2}{\alpha}\right)$ which implies
\begin{equation}
\label{HJKPS9}
f(\alpha) = \beta(\alpha) + \frac{\alpha}{\ln 2}F(\beta(\alpha)) = d_H(\alpha) \, .
\end{equation}

Finally we express the scaling exponents $\zeta_p$ for the structure functions through the
thermodynamic characteristics. The exponent $\zeta_p$ is defined by the scaling
relation
\begin{equation}
\label{structure}
\langle |u(x+l) - u(x)|^p \rangle \, \sim \,  l^{\zeta_p}
\end{equation}
in terms of the integral $u(x)$ of the invariant measure. Let $I_j, \,\, 1\leq
j \leq N= \frac{2}{l}$ be a partition of $[-1,1]$ into intervals of length
$l$. Then
\begin{equation}
\label{structure1}
\langle |u(x+l) - u(x)|^p \rangle  \, = \, \frac{1}{N}\sum_{j=1}^N (\mu_0(I_j))^p \, \sim \,  lN^{D_p(1-p)} \, \sim \,  l^{1+D_p(p-1)}
\end{equation}
which together with (\ref {HJKPS3}) gives
\begin{equation}
\label{structure2}
\zeta_p= 1 + D_p(p-1) = 1-F^{-1}(p\ln2) \, .
\end{equation}
Using (\ref {HJKPS22}) one can also write $\zeta_p$ in the following form:
\begin{equation}
\label{structure3}
\zeta_p= 1 + \inf_{\alpha} \ [\alpha p - f(\alpha)] = 1 + \inf_{\alpha} \ [\alpha p - d_H(\alpha)] \, .
\end{equation}
At $\alpha = \alpha_\star$ the infimum (\ref{structure3}) is attained, 
we can then write the relation between $p$ and $\alpha_\star$ as
\begin{equation}
\label{q_and_alpha} 
 p = d^{\prime}_H (\alpha)|_{\alpha = \alpha_\star} =
 \frac{F(\beta(\alpha_\star))}{\ln 2} \,.
\end{equation}
The scaling exponents for the structure functions are hence obtained as
\begin{equation}
\label{1-beta} 
 \zeta_p = 1 - \beta(\alpha_\star)\,.
\end{equation}

\

We now turn to concrete calculations for the scheme presented above.
It is easy to see that all the thermodynamic functions can be
effectively approximated numerically. The first step is to find 
approximations for the thermodynamic potential 
$U(\epsilon^{(1)}, \epsilon^{(2)}, \dots, \epsilon^{(n)}, \dots)$. We shall use
Markov approximations $U_k(\epsilon^{(1)}, \epsilon^{(2)}, \dots, \epsilon^{(k)})$
which are defined by the following formula:
\begin{equation}
\label{kapprox}
U_k(\epsilon^{(1)}, \epsilon^{(2)}, \dots, \epsilon^{(k)})= \lim _{n \to \infty}
\ln \frac {|\Delta^{(n)}_{1,0,\dots,0,\epsilon^{(k)},\epsilon^{(k-1)}, \dots, \epsilon^{(1)}}|}
{|\Delta^{(n-1)}_{1,0,\dots,0,\epsilon^{(k)},\epsilon^{(k-1)}, \dots, \epsilon^{(2)}}|}.
\end{equation}
It was shown in \cite{vsk84} that the limit in (\ref{kapprox}) exists and
\begin{equation}
\label{kapprox1}
|U(\epsilon^{(1)}, \epsilon^{(2)}, \dots, \epsilon^{(k)}, \epsilon^{(k+1)}, \dots) - U_k(\epsilon^{(1)}, \epsilon^{(2)}, \dots, \epsilon^{(k)})| \leq C(2\gamma)^k \, .
\end{equation}
Using the approximate potential $U_k$ we can construct a Markov approximation for the transfer-matrix
operator ${\cal L}(\beta)$. Namely, we define a finite dimensional linear operator
${\cal L}_k(\beta)$:
\begin{equation}
\label{Lk}
{\cal L}_k(\beta)h(\epsilon^{(1)}, \epsilon^{(2)}, \dots, \epsilon^{(k)}) = \sum_{\epsilon^{(0)} = 0,1}\exp[\beta U_{k+1}(\epsilon^{(0)},\epsilon^{(1)}, \epsilon^{(2)}, \dots, \epsilon^{(k)})]h(\epsilon^{(0)},\epsilon^{(1)}, \epsilon^{(2)}, \dots, \epsilon^{(k-1)}) \, .
\end{equation}
In the matrix representation the operator ${\cal L}_k(\beta)$ corresponds to a certain
$2^{k}\times2^{k}$ matrix. Denote by $\lambda_K(\beta)$ its largest eigenvalue.
Then
\begin{equation}
\label{kapprox2}
F_k(\beta) = \ln \lambda_k(\beta)
\end{equation}
is a natural approximation for the free energy $F(\beta)$. It follows from
(\ref {kapprox1}) that $F_k(\beta)$ converges to $F(\beta)$ exponentially fast
in $C^{\infty}$ topology. Using $F_k(\beta)$ we can effectively approximate
all the multifractal functions which we discussed above. The corresponding
numerical results are presented in the next section.

\subsection{Numerical calculation of the free energy and the scaling
 exponents}
\label{ss:numericalfreeenergy}

Here we show how to construct the transfer-matrix operator ${\cal L}_k$
starting from $k = 0$ to general $k$.
For $k = 0$, the matrix operator ${\cal L}_0$ is just a scalar. Denoting
$\D{n}{\epsilon_1, \ldots, \epsilon_n}$
by $\Dle{n}{\epsilon_1, \ldots, \epsilon_n}$ for clarity, let us consider 
\begin{eqnarray}
 \exp[U_1(\epsilon^{(0)})]
  =
  \lim_{n \to \infty}
  \frac{|\Dle{n}{1, \overbrace{0, \ldots, 0}^{n - 2}, \epsilon^{(0)}}|}
       {|\Dle{n - 1}{1, \underbrace{0, \ldots, 0}_{n - 2}}|},
\label{e:eU1}       
\end{eqnarray}
whose analytical expression is easy to calculate.
The 0-th order approximation of the free energy is given by
\begin{eqnarray}
\label{twoterms}
 F_0(\beta) = \ln[e^{\beta U_1(0)} + e^{\beta U_1(1)}].
\end{eqnarray}
For $\epsilon^{(0)} = 0$, since $\D{n}{0} = [-\alpha^{-n}, \alpha^{-n}]$
and $\D{n}{1} = [g(\alpha^{-n}), 1]$, we have
\begin{eqnarray}
|\Dle{n}{1, 0, \ldots, 0}| = 1 - g(\alpha^{-n}) = c \alpha^{-2n} + O(\alpha^{-4n}),
\end{eqnarray}
where we use the expansion \eqref{g1} of $g(x)$. Thus $c = 1.527\ldots$.
Therefore, the corresponding component of ${\cal L}_0(\beta = 1)$ is
\begin{eqnarray}
 e^{U_1(0)} = \lim_{n \to \infty}
  \frac{|c \alpha^{-2n}       + O(\alpha^{-4n})|}
       {|c \alpha^{-2(n - 1)} + O(\alpha^{-4(n - 1)})|}
 = \alpha^{-2}.
\label{e:eU10} 
\end{eqnarray}
Before considering the second term in the argument of the logarithm in
\eqref{twoterms}, we recall the relation 
\begin{eqnarray}
 g^{(2^{n})}(\alpha^{-n}x) = (-1)^n \alpha^{-n} g(x),
\label{e:nsim}  
\end{eqnarray}
which played a primary role in the proof of
Theorem 4.1. of Ref.~\cite{vsk84}. 
Since $\Dle{n}{1, 0, \ldots, 0, 1} = \D{n}{1 + 2^{n - 1}}$, in
view of (\ref{e:nsim}), we have
\begin{eqnarray}
 |\Dle{n}{1, 0, \ldots, 0, 1}| 
 &=& |g^{(1 + 2^{n - 1})}(0) - g^{(1 + 2^{n - 1})}(\alpha^{-n})|
 \nonumber \\
&=& |c \alpha^{-2(n - 1)}\{[g(\alpha^{-1})]^2 - 1\} + O(\alpha^{-4(n - 1)})|.
\end{eqnarray}
Thus for $\epsilon^{(0)} = 1$, (\ref{e:eU1}) can be rewritten as
\begin{eqnarray}
 e^{U_1(1)} = \lim_{n \to \infty}
  \frac{|c \alpha^{-2(n - 1)}\{[g(\alpha^{-1})]^2 - 1\} + O(\alpha^{-4(n - 1)})|}
       {|c \alpha^{-2(n - 1)} + O(\alpha^{-4(n - 1)})|}
 = 1 - [g(\alpha^{-1})]^2.
\end{eqnarray}
We now arrive at the expression for the $0$-th order approximation of the free energy:
\begin{eqnarray}
 F_0(\beta) = \ln[\alpha^{-2\beta} + \{1 - [g(\alpha^{-1})]^2\}^{\beta}].
\label{e:f0beta}  
\end{eqnarray}

Next, consider the $k = 1$ approximation to the transfer matrix 
${\cal   L}_1(\beta)$. From \eqref{Lk}, it can be written in  standard matrix 
notation as 
\begin{equation}
\left(
  \begin{array}{cc}
   e^{\beta U_2(0, 0)} & e^{\beta U_2(1, 0)}\\
   e^{\beta U_2(0, 1)} & e^{\beta U_2(1, 1)}
  \end{array}
 \right).
\end{equation} It then follows that the free energy is given by
\begin{eqnarray}
 F_1(\beta)
 &=& \ln \left[e^{\beta U_2(0, 0)} + e^{\beta U_2(1, 1)} +
         \sqrt{(e^{\beta U_2(0, 0)} - e^{\beta U_2(1, 1)})^2
	 + 4 e^{\beta U_2(1, 0)}e^{\beta U_2(0, 1)}}\right] \nonumber \\
&& - \ln 2.
\end{eqnarray}
The four components of the transfer matrix require the evaluation of suitable
exponential terms, expressible by \eqref{kapprox}, from 
\begin{eqnarray}
 \exp[U_2(\epsilon^{(0)}, \epsilon^{(1)})] = \lim_{n \to \infty}
  \frac{|\Dle{n} {1, \overbrace{0, \ldots, 0}^{n - 3}, \epsilon^{(1)},
  \epsilon^{(0)}}|} {|\Dle{n - 1}{1, \underbrace{0, \ldots, 0}_{n - 3},
  \epsilon^{(1)}}|}.
\end{eqnarray}
Each of these terms is calculated in the same manner as for the $k = 0$ case:
\begin{eqnarray}
 e^{U_2(0, 0)} &=&
 \lim_{n \to \infty}
 \frac{|\Dle{n}    {1, 0, \dots, 0}|}
      {|\Dle{n - 1}{1, 0, \dots, 0}|}
 = \alpha^{-2},\\
 e^{U_2(1, 0)} &=&
 \lim_{n \to \infty}
 \frac{|\Dle{n}    {1, 0, \dots, 0, 1}|}
      {|\Dle{n - 1}{1, 0, \dots, 0}   |} = [g(0)]^2 - [g(\alpha^{-1})]^2 \nonumber\\
  &=&  1 - [g(\alpha^{-1})]^2,\\
 e^{U_2(0, 1)} &=&
 \lim_{n \to \infty}
 \frac{|\Dle{n}    {1, 0, \dots, 0, 1, 0}|}
      {|\Dle{n - 1}{1, 0, \dots, 0, 1}   |} \nonumber \\
 &=&
 \frac{|[g(0)]^2 - [g(\alpha^{-2})]^2|}
      {|[g(0)]^2 - [g(\alpha^{-1})]^2|}
 =       
 \frac{1 - [g(\alpha^{-2})]^2}
      {1 - [g(\alpha^{-1})]^2},\\ 
 e^{U_2(1, 1)} &=&
 \lim_{n \to \infty}
 \frac{|\Dle{n}    {1, 0, \dots, 0, 1, 1}|}
      {|\Dle{n - 1}{1, 0, \dots, 0, 1}   |} \nonumber \\
&=&
 \frac{|[g^{(3)}(0)]^2 - [g^{(3)}(\alpha^{-2})]^2|}
      {|[g(0)]^2 - [g(\alpha^{-1})]^2|}.      
\end{eqnarray}

For $k = 2$, we just write down the transfer matrix ${\cal L}_2(\beta)$:
\begin{equation}
 \left(
  \begin{array}{cccc}
   e^{\beta U_3(0, 0, 0)} & 0                      & e^{\beta U_3(1, 0, 0)} & 0                      \\
   e^{\beta U_3(0, 0, 1)} & 0                      & e^{\beta U_3(1, 0, 1)} & 0                      \\
   0                      & e^{\beta U_3(0, 1, 0)} & 0                      & e^{\beta U_3(1, 1, 0)} \\
   0                      & e^{\beta U_3(0, 1, 1)} & 0                      & e^{\beta U_3(1, 1, 1)}
  \end{array}
 \right).
\end{equation}

Now we are in a position to calculate the transfer matrix
${\cal L}_k(\beta)$ for general $k$.
Let us consider the component for $\beta = 1$:
\begin{eqnarray}
 \exp[U_{k + 1}(\epsilon^{(0)}, \epsilon^{(1)}, \ldots, \epsilon^{(k)})]
&=&
\lim_{n \to \infty}
\frac{|\Dle{n}{1, \overbrace{0, \ldots, 0}^{n - k - 2}, \epsilon^{(k)}, \ldots, \epsilon^{(1)}, \epsilon^{(0)}}|}
     {|\Dle{n - 1}{1, \underbrace{0, \ldots, 0}_{n - k - 2}, \epsilon^{(k)}, \ldots, \epsilon^{(1)}}|} \nonumber \\
&=&
\lim_{n \to \infty} 
\frac{|g(\Dle{n}{\overbrace{0, \ldots, 0}^{n - k - 2}, \epsilon^{(k)}, \ldots, \epsilon^{(1)}, \epsilon^{(0)}})|}
     {|g(\Dle{n - 1}{\underbrace{0, \ldots, 0}_{n - k - 2}, \epsilon^{(k)}, \ldots, \epsilon^{(1)}})|}.
\label{e:compo}     
\end{eqnarray}
Hence it is enough to calculate $|\Dle{n}{0, \ldots, 0, \epsilon^{(k)},
\ldots, \epsilon^{(1)}, \epsilon^{(0)}}|$.  For this we use the following 
relation:
\begin{eqnarray}
&&|\Dle{n}{\underbrace{0, \ldots, 0}_{n - k - 2}, \epsilon^{(k)}, \ldots, \epsilon^{(1)}, \epsilon^{(0)}}|   
\nonumber \\
&& \hspace{-3cm}
 = |g^{(\epsilon^{(k)}2^{n - k - 1} + \ldots + \epsilon^{(1)}2^{n - 2} + \epsilon^{(0)}2^{n - 1} )}(0)
 -
 g^{(\epsilon^{(k)}2^{n - k - 1} + \ldots + \epsilon^{(1)}2^{n - 2} + \epsilon^{(0)}2^{n - 1} )}(\alpha^{-n})|
\nonumber  \\
&=&
  \alpha^{-(n - k)}|\Dle{k + 1}{\epsilon^{(k)}, \ldots, \epsilon^{(1)}, \epsilon^{(0)}}|.
\label{e:similarity}
\end{eqnarray}
By (\ref{e:similarity}) the numerator in (\ref{e:compo}) is given by
\begin{eqnarray}
 |\D{n}{1 + j\cdot 2^{n - k - 1}}|
  &=& |\{1 - c\, \alpha^{-2(n - k - 1)}[g^{(j)}(0)]^2\}
     -
     \{1 - c\, \alpha^{-2(n - k - 1)}[g^{(j)}(\alpha^{-(k + 1)})]^2\} + O(\alpha^{-4(n - k)})|\} \nonumber \\
  &=& c \, \alpha^{-2(n - k - 1)}|[g^{(j)}(0)]^2 - [g^{(j)}(\alpha^{-(k + 1)})]^2| + O(\alpha^{-4(n - k)}).
\end{eqnarray}
Therefore the component (\ref{e:compo}) is expressed as
\begin{eqnarray}
 \exp[U_{k + 1}(\epsilon^{(0)}, \epsilon^{(1)}, \ldots, \epsilon^{(k)})]
  =
 \left|
  \frac{[g^{(j)}(0)]^2  - [g^{(j)}(\alpha^{-(k + 1)})]^2}
       {[g^{(j')}(0)]^2 - [g^{(j')}(\alpha^{-k})]^2}
 \right|,
\label{e:concretecompo} 
\end{eqnarray}
where 
$j = \epsilon^{(k)} + \epsilon^{(k - 1)} 2 + \ldots +
\epsilon^{(1)} 2^{k - 1} + \epsilon^{(0)} 2^{k}$ and
$j' = \epsilon^{(k)} + \epsilon^{(k - 1)} 2 + \ldots +
\epsilon^{(1)} 2^{k - 1}$.
In other words, if the coordinates of the end points of the intervals
are given by $\D{k + 1}{j} = [a_j^{(k + 1)},\, b_j^{(k + 1)}]$
and $\D{k}{j'} = [a_{j'}^{(k)},\, b_{j'}^{(k)}]$,
the component (\ref{e:compo}) can be rewritten as
\begin{eqnarray}
\exp[U_{k + 1}(\epsilon^{(0)}, \epsilon^{(1)}, \ldots, \epsilon^{(k)})]
 =
 \left|
  \frac{\left(a_j^{(k + 1)}\right)^2  - \left(b_j^{(k + 1)}\right)^2}
       {\left(a_{j'}^{(k)}   \right)^2  - \left(b_{j'}^{(k)}   \right)^2}
 \right|.
 \label{e:ab}
\end{eqnarray}
If the transfer matrix is written as an ordinary $2^{k}\times 2^{k}$
matrix, it is easily found that the row and columns indices are given by
$\mbox{(row, column)} = (p, q)$, where
\begin{eqnarray}
({\rm row}) \quad   p &=&  1 + \epsilon^{(k)} + \epsilon^{(k - 1)}2 + \ldots + \epsilon^{(1)} 2^{k- 1},\\
({\rm column}) \quad q &=& 1 + \epsilon^{(k - 1)} + \epsilon^{(k - 2)}2 + \ldots + \epsilon^{(0)} 2^{k- 1}.
\end{eqnarray}

\begin{figure}
\iffigs 
\centerline{%
\includegraphics[scale=0.8]{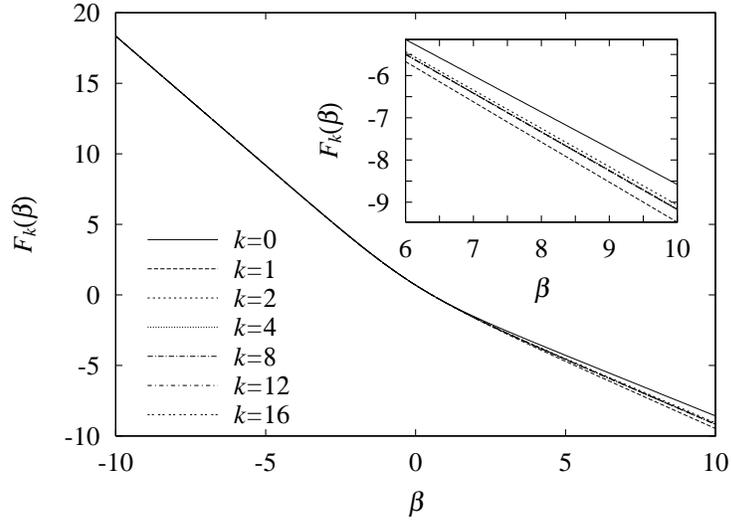}
}
\else\drawing 65 10 {free energy}
\fi
\caption{\label{f:free-energy}%
The $k$-th order approximation to the free energy $F_k(\beta)$.
Inset: Enlargement of the region $6 \le \beta \le 10$; when increasing
$k$, alternate convergence  is observed.
}
\end{figure}

\begin{figure}
\iffigs 
\centerline{%
\includegraphics[scale=0.75]{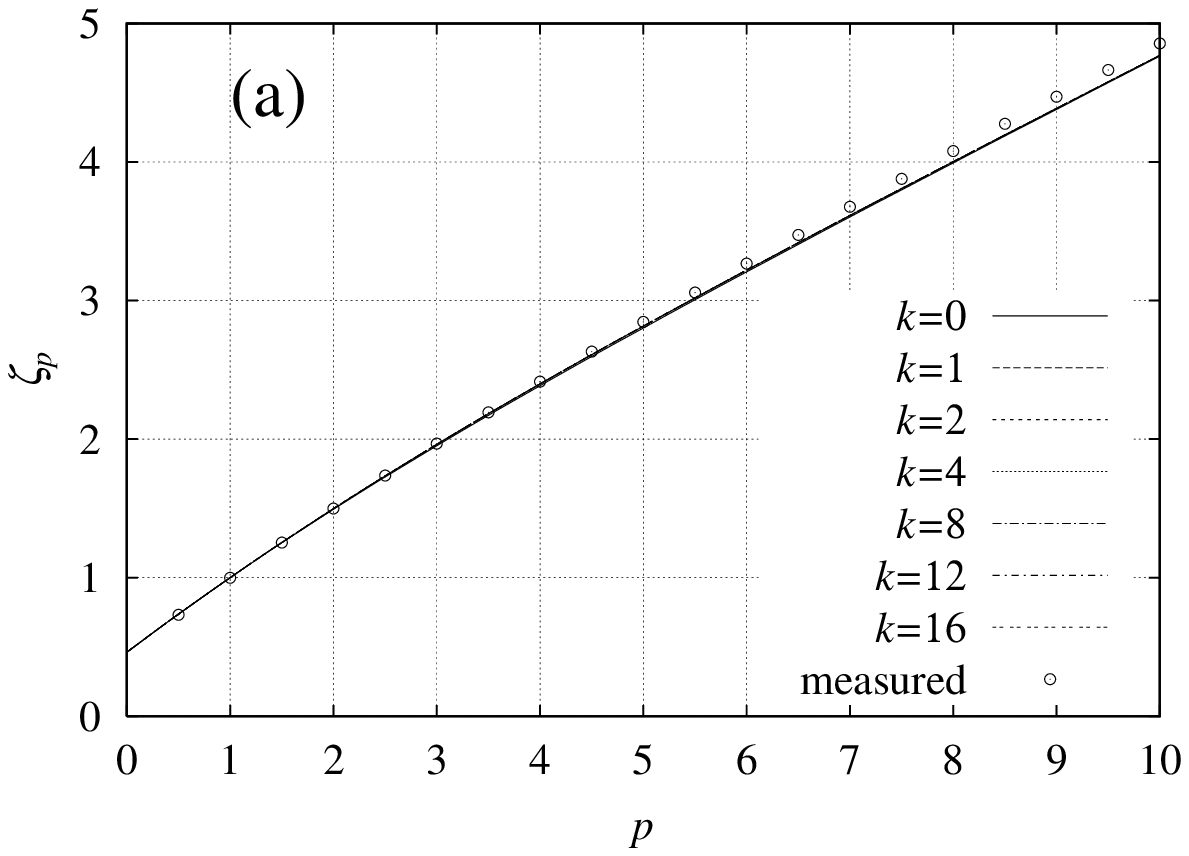} 
\includegraphics[scale=0.75]{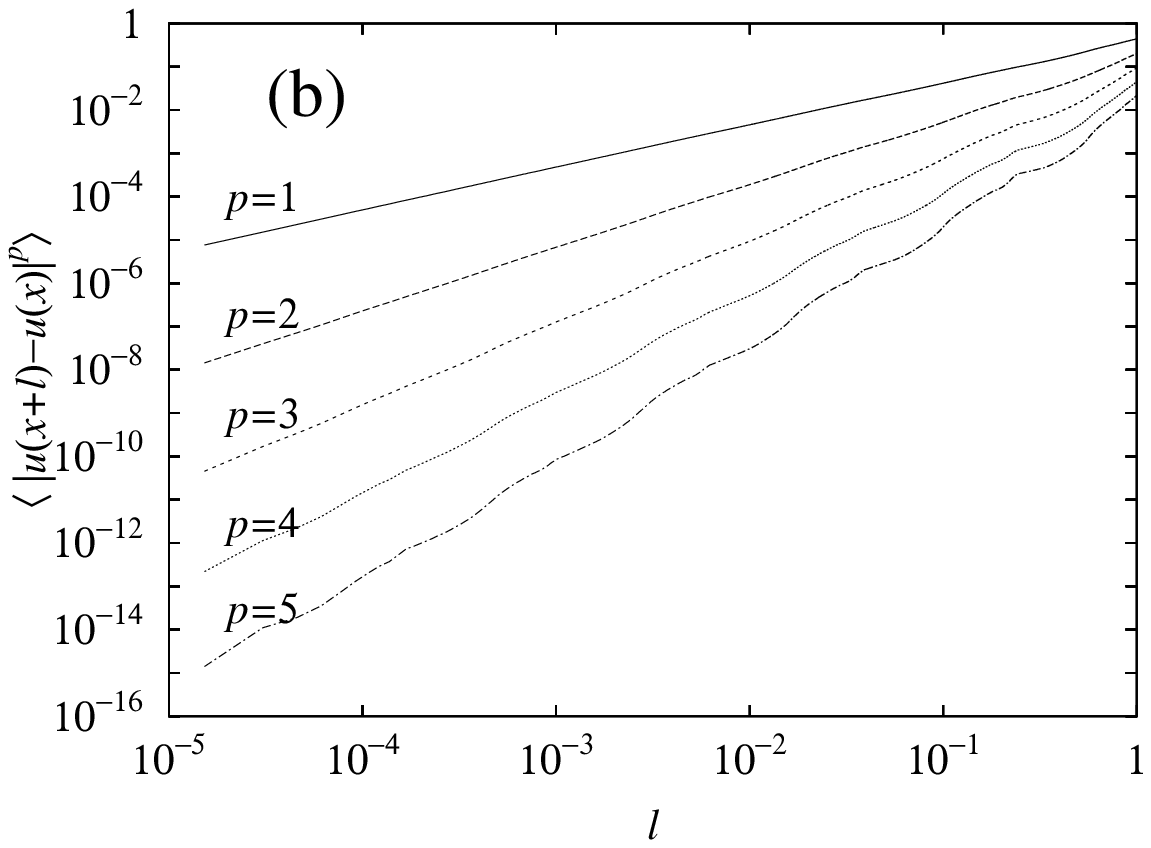}
}
\else\drawing 65 10 {feigen: structure func, zetap}
\fi
\caption{\label{f:feigen-zetap}%
(a) Scaling exponents $\zeta_p$ of structure functions obtained by two
 methods. Open circles: data obtained by a least-square fit of the slopes of
 the directly measured structure functions shown in (b).  Lines: data
 calculated from the free energy $F_k(\beta)$ using \eqref{structure2} (curves
 for different $k$ are essentially indistinguishable).
}
\end{figure}

For numerical calculation of the functions $F_k(\beta)$ and $\zeta_p$
associated to $u(x)$, we use the expansion \eqref{g1} for $g(x)$, as given in
Ref.~\cite{l82}.  We include terms up to $x^{32}$. Numerical calculation is
done with standard double precision (15 significant digits).  For obtaining
the largest eigenvalue, we use the power method for matrices \cite{gv96}.  The
multiplication of the matrix is stopped when the relative error of the most
dominant eigenvalue becomes less than $10^{-13}$, thus giving $10^{-13}$
absolute error on the $F(\beta)$ function.
An alternative approach, also based on
the thermodynamic formalism and yielding 10-digit accuracy, may be found in
Ref.~\cite{k89}. Approximate free energies $F_k(\beta)$, with $k$ up to $16$,
calculated by the transfer-matrix method are shown in \figref{f:free-energy}.
We note that the $k = 0$ approximation (\ref{e:f0beta}) already gives a
reasonable estimate.  The discrepancy of the free energy between various
orders of approximations is visible at large $\beta$.  However the $\beta > 0$
region is irrelevant as far as $\zeta_p$ for positive $p$ is concerned (see
Eq. (\ref{1-beta})).  The corresponding $\zeta_p \, (p \ge 0)$ are calculated
from the $F_k(\beta)$ by (\ref{structure2}) for different values of $k$; the
results, which hardly depend on $k$, are shown in \figref{f:feigen-zetap}(a).
We also determined the structure functions of $u(x)$ with $2^{17}$ uniform
bins in $[-1, 1]$; they are plotted in \figref{f:feigen-zetap}(b). The
exponents $\zeta_p$ are then obtained by a least square fit of the
structure functions over the range $2^{-16} \sim 1.5\times10^{-6} \le r \le
1$.  With this number of bins, the quality of the fit begins to somewhat
deteriorate beyond $p=4$, but otherwise there is rather good agreement
between the two methods of determining $\zeta_p$. Note that the
``$y$-intercept'' of the graph of $\zeta_p$, namely $\zeta_0$, which is the
codimension of the support of the invariant measure $\mu_0(dx)$, is 
positive and its numerical value is slightly under one half \cite{vsk84}. 
This will be important in the sequel.
\section{Fractional derivatives for  the Feigenbaum
 attractor}
\label{s:fraclap}
 
\subsection{Phenomenology for multifractality and fractional derivatives}
\label{ss:fraclap_pheno}

In this section we briefly recall the phenomenological approach to
multifractality via fractional derivatives \cite{fm02} and adapt it to a
multifractal measure. We therefore work, not with the measure $\mu_0(dx)$
itself, but with its integral $u(x)$. Singularity exponents $\alpha$ may be
viewed as local H\"{o}lder exponents of $u(x)$, i.e., $|u(x + l) - u(x)|
\propto |l|^{\alpha}$ for $(l \to 0)$. We turn to fractional derivatives of
order $a$ defined, as in Ref.~\cite{fm02}, as the multiplication in the
Fourier space by $\hat{u}(k)$ by $|k|^{a}$ (see Ref.~\cite{fm02} for precise
definition). An isolated non-oscillatory 
singularity with  exponent $\alpha$
at a point $x$ implies \begin{equation}
 |D^{a} u(y)| \sim |y - x|^{\alpha - a} \quad (y \to x).
\label{e:neighbour}  
\end{equation}
If $\alpha - a < 0$, as we shall assume hereafter, the exponent is
negative,  the fractional derivative can become arbitrarily large and
thus contributes to the tail-behavior of the probability.
A key assumption in the phenomenology is that this argument can be carried
over to non-isolated multifractal singularities, provided we take
all types of singularities into account. For the Feigenbaum invariant
measure, we know the Hausdorff dimension  $d_H(\alpha)$ of the set of points 
having a singularity with exponent $\alpha$. Assuming that we can also
use $d_H(\alpha)$ as  a covering dimension, we can express the probability
to have a singularity of exponent $\alpha$ contributing a
fractional derivative of order $a$ which exceeds (in absolute value)
a given large value $\xi$, that is we require
\begin{equation}
 |y - x| < \xi^{-\frac{1}{a - \alpha}}.
\label{e:y-x}  
\end{equation}
In terms of  the codimension of the set ${\cal I}_{\alpha}$, the probability
to satisfy \eqref{e:y-x} is written as
\begin{equation}
{\rm  Prob}\{|D^{a} u| > \xi\} \propto |y - x|^{d - d_H(\alpha)} 
\propto
\xi^{-\frac{d - d_H(\alpha)}{a - \alpha}}.
\label{e:prob_single} 
\end{equation}
Here $d$ is the spatial dimension ($d = 1$). 
Taking now into account  the singularities with all possible exponents 
$\alpha$, the tail of the cumulative probability of the fractional
derivative of order $a$ is given, to the  leading order, by the following
power law
\begin{eqnarray}
 {\rm  Prob}\{|D^{a} u| > \xi\} \propto \xi^{-p_\star}\, , \quad \xi \to \infty,
\label{e:cprob}\\
 p_{\star} = \inf_{\alpha < a} \frac{d - d_H(\alpha)}{a - \alpha}.
\label{e:inf} 
\end{eqnarray}
An easy calculation shows that $\alpha_\star$ corresponding to the infimum in \eqref{e:inf}
satisfies 
\begin{equation}
\label{alpha_star}
\alpha_\star=\alpha_\star(a)=a + \frac{d_H(\alpha_\star) -d}{d'_H(\alpha_\star)},
\end{equation}
which immediately gives $p_\star=d'_H(\alpha_\star)$. On the other hand, we
know that 
\begin{equation}
\zeta_p=\inf_\alpha(p\alpha +d -d_h(\alpha)).
\end{equation}
Here  the infimum is given by an $\alpha$ satisfying the very same relation 
$p_\star=d'_H(\alpha_\star)$. Hence,
\begin{equation}
\zeta_{p_\star}=p_\star\alpha_\star + d - d_H(\alpha_\star).
\end{equation}
Using (\ref{alpha_star}), we get
\begin{equation}
\zeta_{p_\star}=p_\star\left(a + \frac{d_H(\alpha_\star) -d}{d'_H(\alpha_\star)}\right)+d 
-d_H(\alpha_\star)= p_\star a,
\end{equation}
where the second relation follows from $p_\star=d'_H(\alpha_\star)$.
The geometrical interpretation of this equation is that the (negative)
exponent of the power-law tail for the fractional derivative of order $a$ is 
the $p$-value of the intersection of the graph of $\zeta_p$ and of a straight
line of slope $a$ through the origin. 

As shown in Ref.~\cite{fm02},  in the presence of the finite range of scaling,
the  power-law tail  (\ref{e:cprob}) emerges only if the multifractality
is sufficiently strong. This strength is given  by the 
multifractality parameter $C(a)$, a measure of how strongly the data depart
from being self similar (which would imply $\zeta_p \propto p$): 
\begin{eqnarray}
C(a) \equiv a - \alpha_\star = \frac{\zeta_\pstar}{p_\star} -
\left.\frac{d\zeta_p}{dp}\right|_{p = \pstar}\, ,
\label{e:defC}  
\end{eqnarray}
where $\alpha_\star \equiv d\zeta_p/dp|_{p
= p_\star}$. It was shown  that observability of the power-law 
requires a sufficiently large value for the product $n C(a)$, where $n$ is the
number of octaves over which the data present multifractal  scaling.
In practice it was found in Ref.~\cite{fm02} that
\begin{equation}
n C(a) \ge 10.
\label{e:criterion} 
\end{equation}
For example, fully-developed turbulence velocity data \cite{agha84}
have typical $C$ values of the order of $1/30$, thereby requiring
a monstrous inertial range of about  300~octaves for observability
of power-law tails. As we shall see, the situation
is much more  favorable for the Feigenbaum invariant  measure.

 Before turning to numerical questions, we comment on an issue raised by
an anonymous referee who worried about the nonlocal character of the
fractional derivative and wrote in essence that our approach makes
sense, strictly speaking, only for (statistically) translationally
invariant in space systems: otherwise, if the system consists of
components whose ``fractal properties'' are rather different the
results will be smeared out. Our feeling about such matters is
summarized as follows. First one can observe that, of course, the
attractor for the Feigenbaum map is not homogeneous
(translation-invariant) but after zooming in it becomes increasingly
so; the fractional derivative is not a local operator but the tail of
its PDF is likely to be dominated by strongly localized
events. Second, a more technical observation.  The idea of the
multi-fractal analysis is based on the fact that the dynamics of a
system determines a variety of scales.  It is important that these
scales do not depend on a particular place in the phase space. On the
contrary, they are present and ``interact'' with each other
everywhere. In the case of the Feigenbaum attractor the scales depend
on a symbolic location in a system of partitions.  In physical
systems, like homogeneous turbulence, such partitions are difficult to
define rigorously. However, the invariance with respect to the space
coordinate is still present and forms a basis for applicability of the
multifractal calculus.

\subsection{Numerical analysis of fractional derivatives}
\label{ss:fraclap_numerics}

The phenomenological arguments presented in the previous section suggest that
we should find power-law tails in the cumulative probability for fractional
derivatives of $u(x)$ for suitable orders $a$. Inspection of
\figref{f:feigen-zetap} indicates that $a$ should be between the minimum slope
of the graph and unity. The value $a=1$ is of course not a
fractional order but, as we shall see, it is associated with a power-law tail
of exponent minus one.\footnote{This is a consequence of $\zeta_1 = 1 -
F^{-1}(\ln 2) = 1$.  (Putting $\beta = 0$ into the partition function
$Z_k(\beta) \sim \sum_{j = 0}^{2^k - 1} |\D{k}{j}|^\beta$, we have $F_k(0) =
\ln Z_k(0)/k = \ln 2$.)}
The minimum slope can be easily found. Indeed, $F(\beta)$ takes large values
when $\beta$ is large negative. In this case the main contribution to $F(\beta)$comes from the shortest interval of the partition with the length of the order
of $\gamma^{2n}$. Hence, $F(\beta)/|\beta| \to -2\ln\gamma$ in the limit 
$\beta \to -\infty$. This gives the following lower bound of the
differentiation order:
\begin{equation}
\lim_{p \to \infty}\frac{\zeta_p}{p}=\lim_{p \to 
\infty}\frac{1-F^{-1}(p\ln2)}{p}=-\frac{\ln 2}{2\ln \gamma}\approx 0.3777.
\end{equation}
We have already observed that, because the $\zeta_p$
graph does not pass through the origin, substantial values can be expected for
the multifractality parameter $C(a)$.  The actual values of $C$, associated to
values of $p$ ranging from $1$ to 3 by increments of $0.5$ are shown in
Table~\ref{t:fzetap}, together with the number $n$ of scaling octaves needed
determined by $n C \approx 10$ (cf.\ Eq.~\eqref{e:criterion}).  
\begin{table}
\caption{\label{t:fzetap} For the Feigenbaum invariant measure we
show the scaling exponents $\zeta_p$, the corresponding inverse temperature
$\beta$'s, the multifractality parameter $C$ and the  number of scaling octaves
needed.}
\begin{center} 
 \begin{tabular}{ccccccc}
  \hline
 $p$ & $\zeta_p$          & $\beta$        & $C$  & number of octaves needed  \\ \hline
 1.0 &  1.0               &  0.0           & 0.48 & 21 \\ \hline
 1.5 &  1.2540292658      & -2.54029265895 & 0.34 & 29 \\ \hline
 2.0 &  1.4985620106      & -4.98562010659 & 0.28 & 36 \\ \hline
 2.5 &  1.7344372955      & -7.34437295506 & 0.24 & 42 \\ \hline
 3.0 &  1.9625763533      & -9.62576353333 & 0.22 & 45 \\ \hline
 \end{tabular}
\end{center} 
\end{table}
In practice, on a 32 bit machine, we are limited to about 25 octaves of
dynamical range in resolution over the interval $[-1,\, 1]$. This should be
enough to observe power-law tails.

\begin{figure}
\iffigs
\centerline{%
\includegraphics[scale=0.72]{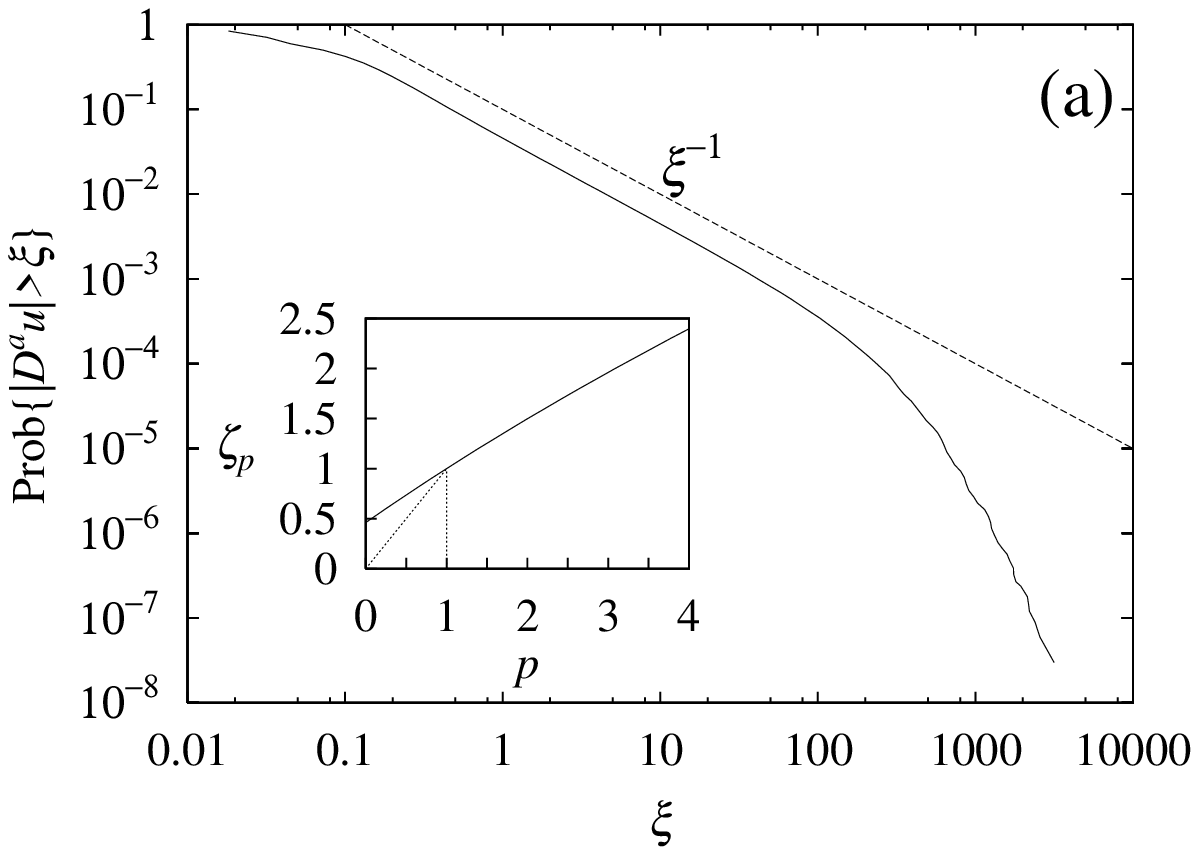}
}
\centerline{%
\includegraphics[scale=0.72]{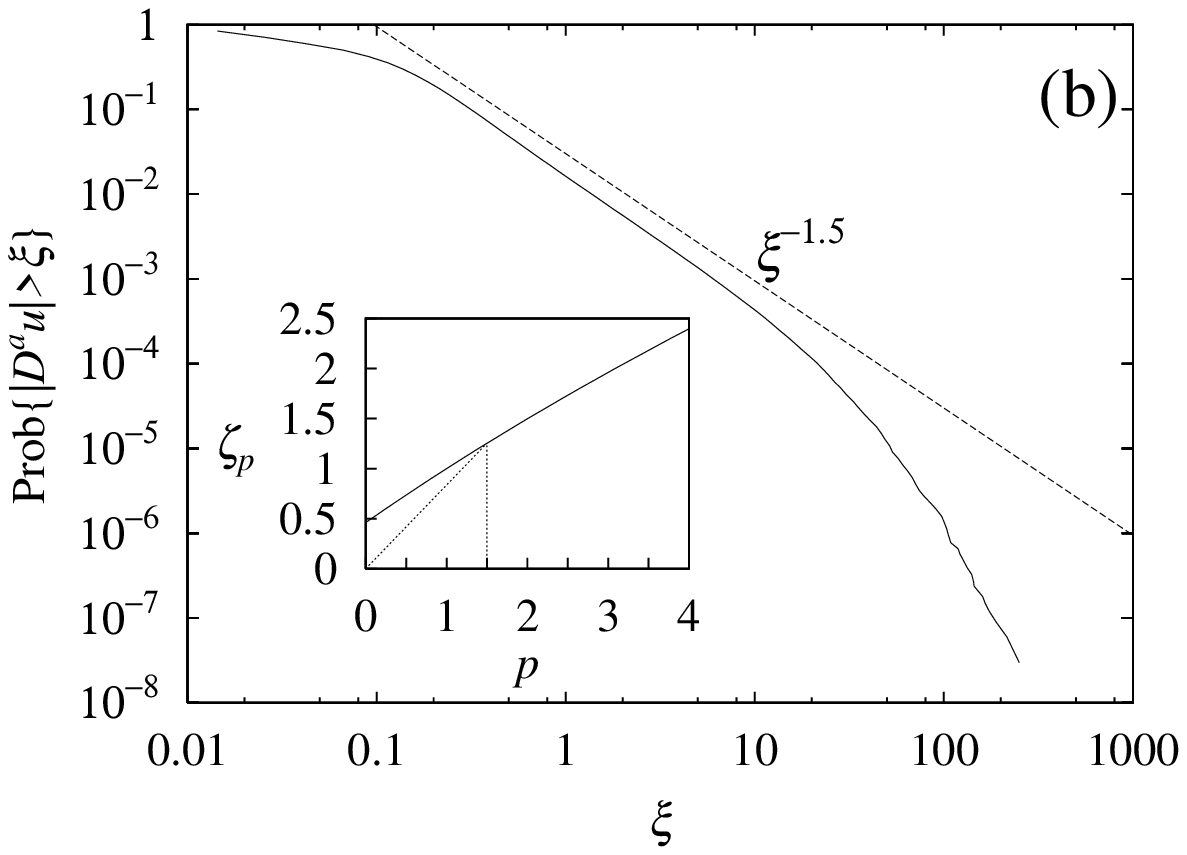} 
\includegraphics[scale=0.72]{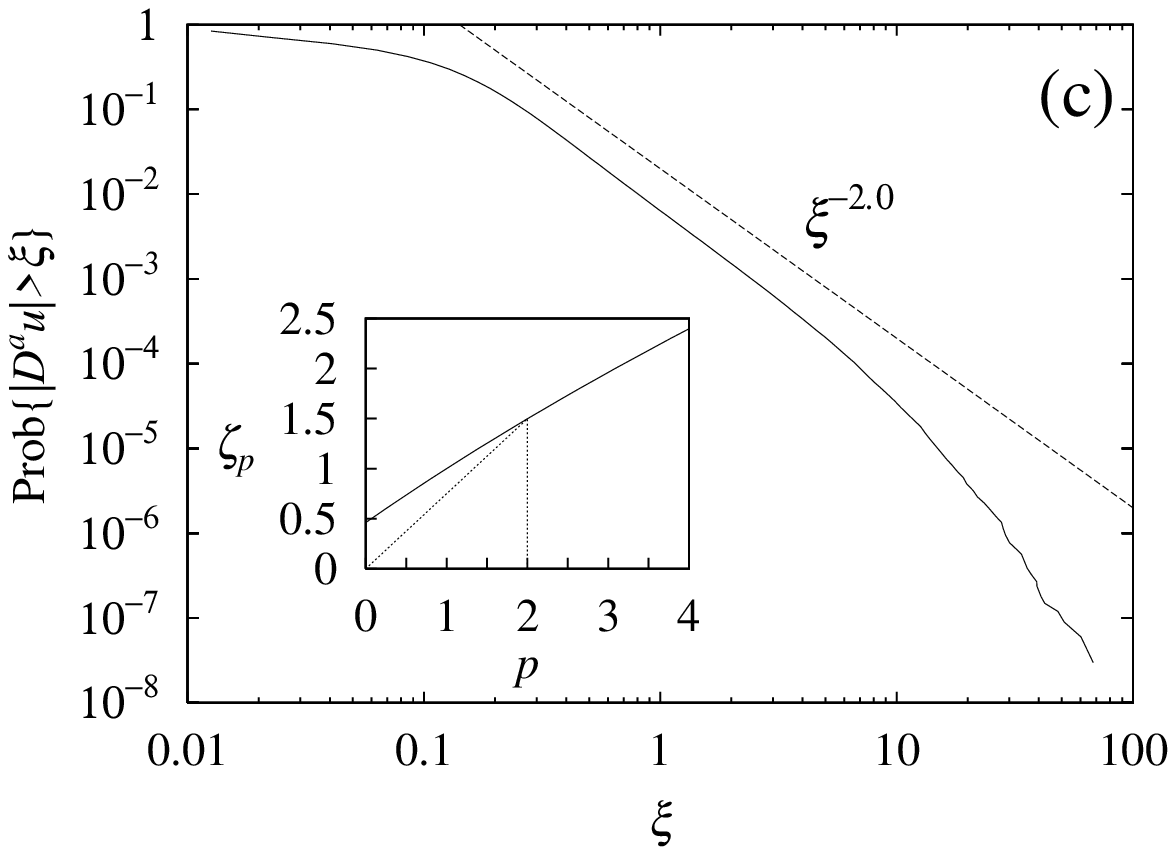}
}
\centerline{%
\includegraphics[scale=0.72]{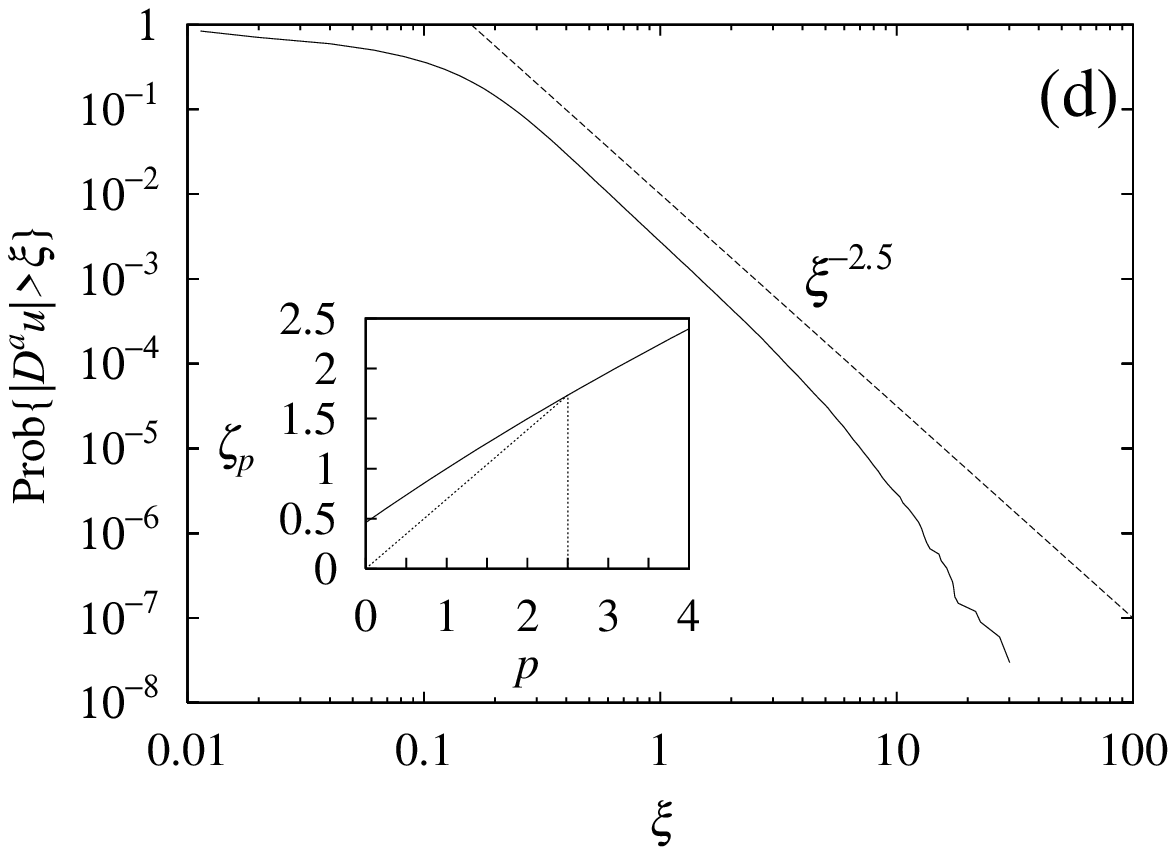} 
\includegraphics[scale=0.72]{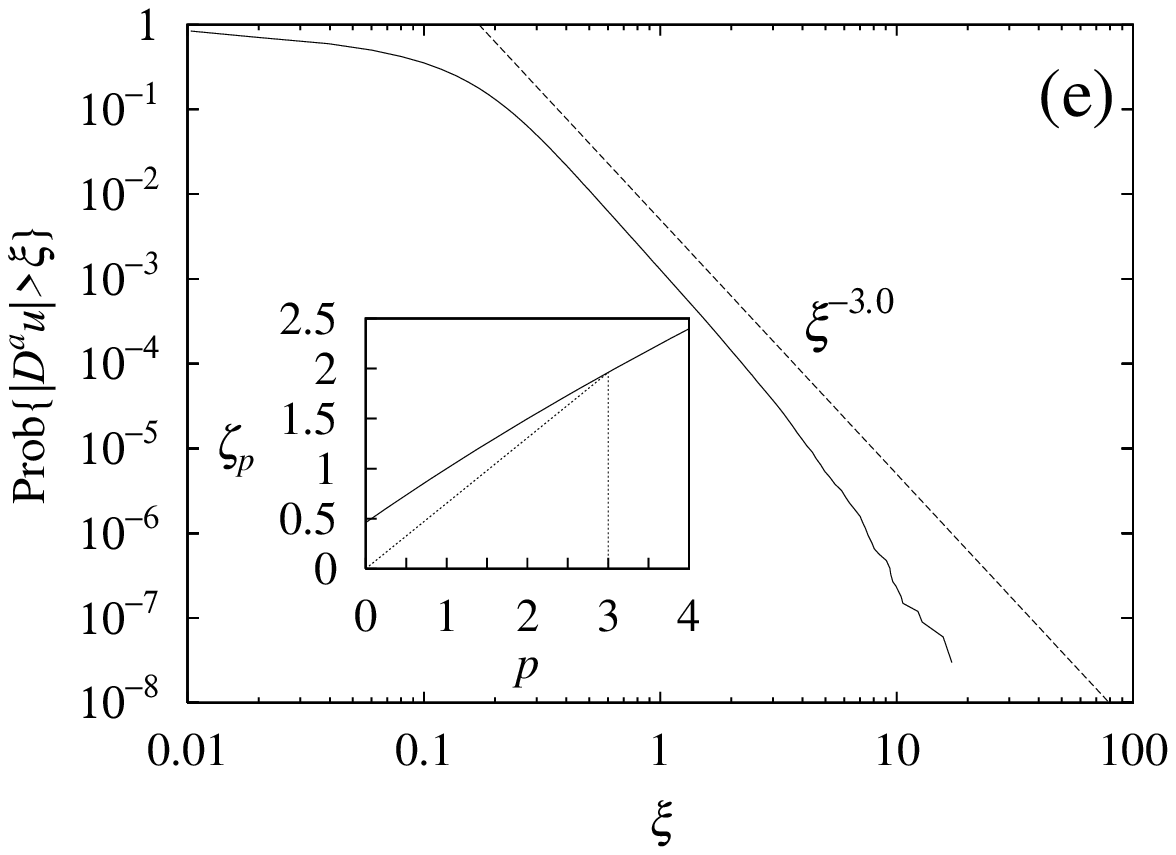}
}
\else\drawing 65 10 {feigen: cumulative probability}
\fi
\caption{\label{f:feigen-cump}%
Cumulative probabilities of absolute values of fractional derivatives of
 various orders $a = \zeta_\pstar/\pstar$ for 
the Feigenbaum invariant measure.
 Each function displays a  power-law tail with an exponent
 fairly close to the predicted value $-\pstar$.
 Insets: corresponding $\zeta_p$ graphs.
(a) $a = 1.0             ~(\pstar = 1.0)$. 
(b) $a = 0.83 ~(\pstar = 1.5)$.
(c) $a = 0.74 ~(\pstar = 2.0)$.
(d) $a = 0.69 ~(\pstar = 2.5)$.
(e) $a = 0.65 ~(\pstar = 3.0)$.
}
\end{figure}
Indeed, \figref{f:feigen-cump} shows five instances of cumulative
probabilities of fractional derivatives with power-law tails, corresponding to
the values of the exponent $p$ listed in Table~\ref{t:fzetap}.  The
corresponding order of differentiation $a$ ranges between $0.65$ and $1$.\,\footnote{As predicted by the theory, when $a$ is too small, e.g. for $a=0$,
no power-law tail is observed.}
Since the function $u(x)$ which we are analyzing is not periodic, 
we resort to the Hann windowing technique employed previously
in Ref.~\cite{fm02} (Section 13.4). Also, we use 
rank ordering to avoid binning. 

The power-law behavior observed is consistent with the phenomenological theory
presented in Section~\ref{ss:fraclap_pheno}, the residual discrepancies being
due to the resolution of $2^{25}$ bins.

\section{Concluding remarks}
\label{s:concl}

We have found solid numerical evidence for the presence of power-law
tails in the cumulative distribution of fractional derivatives
for the integral  $u(x)$ of the invariant measure of the Feigenbaum
map. Furthermore the exponents measured are consistent
with those predicted by phenomenological arguments from the
spectrum of singularities. Since we have a fairly deep understanding
of the structure of the attractor, thanks in particular to the 
thermodynamic formalism, a reasonable goal may be to actually prove
the results. The main difficulty is that the operation of fractional
derivative is non-local. However, we believe that a rigorous analysis here is
still possible due to the quite simple spectral structure of the
dynamical system corresponding to the Feigenbaum attractor.

\section*{Acknowledgments}
We are grateful to Rahul Pandit for useful remarks. Computational resources
were provided by the Yukawa Institute (Kyoto). This research was supported by
the European Union under contract HPRN-CT-2000-00162 and by the Indo-French
Centre for the Promotion of Advanced Research (IFCPAR~2404-2).


\begin{thebibliography}{99}
\bibitem{fm02}
U.~Frisch and T.~Matsumoto,
On multifractality and fractional derivatives,
{\it J.~Stat.~Phys.} {\bf 108}:1181--1202 (2002).

\bibitem{benzi93}
R.~Benzi, L.~Biferale, A.~Crisanti, G.~Paladin, M.~Vergassola and A.~Vulpiani,
A random process for the construction of multiaffine fields,
{\it Physica D} {\bf 65}:163--171 (1993).
    
\bibitem{kesten73}
H.~Kesten,
Random difference equations and renewal theory for products of random matrices,
{\it Acta. Math.} {\bf 131}:207--248 (1973).
 
\bibitem{f78}
M.J.~Feigenbaum,
Quantitative universality for a class of nonlinear 
transformations, {\it J.\ Stat. Phys.} {\bf 19}:25--52 (1978).

\bibitem{f80}
M.J.~Feigenbaum, 
The transition to aperiodic behavior in turbulent systems, 
{\it Comm.\ Math.\ Phys.} {\bf 77}:65--86 (1980).
	
	
\bibitem{vsk84}
E.B.~Vul, Ya.G.~Sinai and K.M.~Khanin,
Feigenbaum universality and the thermodynamic formalism,
{\it Russian Math. Surveys} {\bf 39}:1--40 (1984).

\bibitem{pf85}
G.~Parisi and U.~Frisch, 
On the singularity structure of fully developed turbulence,
in {\it Turbulence and Predictability in Geophysical Fluid
Dynamics}, Proceedings of International School of Physics 'Enrico Fermi', Jun. 14--24 1983, Varenna, Italy,
M.~Ghil, R.~Benzi and G.~Parisi, eds.,
pp.~84--87, North Holland (1985).

\bibitem{bppv84}
R.~Benzi, G.~Paladin, G.~Parisi and A.~Vulpiani,
On the multifractal nature of fully developed turbulence and chaotic systems
{\it J.~Phys.~A}{\bf 17}:3521--3531 (1984).	

	
\bibitem{HJKPS}
T.C.~Halsey, M.H.~Jensen, L.P.~Kadanoff, I.~Procaccia and B.I.~Shraiman,
Fractal measures and their singularities: the characterization of
strange sets,
{\it Phys. Rev. A} {\bf 33}:1141--1151 (1986).
	
\bibitem{l82}
O.E.~Lanford,
A computer-assisted proof of the Feigenbaum conjectures,
{\it Bull. Ame. Math. Soc.} {\bf 6}:427--434 (1982).

	
\bibitem{frisch95}
U.~Frisch,
{\it Turbulence,  the Legacy of A. N. Kolmogorov},
Cambridge University Press, Cambridge (1995).


\bibitem{cla87}
P.~Collet, J.L.~Lebowitz and A.~Porzio,
The dimension spectrum of some dynamical systems,
{\it J. Stat. Phys.} {\bf 47}:609--644 (1987).	
	
\bibitem{gv96}
G.H.~Golub and C.F.~Van Loan,
{\it Matrix Computations} 3rd edition,
The Johns Hopkins University Press, Baltimore (1996),
Section 7.3.


\bibitem{k89}
Z.~Kov\'acs, Universal $f(\alpha)$ spectrum as an eigenvalue,
{\it J.\ Phys.} A {\bf 22}:5161--5165 (1989).

	
\bibitem{agha84}
F.~Anselmet, Y.~Gagne, E.J.~Hopfinger and R.A.~Antonia,
High-order velocity structure functions in turbulent shear flow,
{\it J. Fluid Mech.} {\bf 140}:63--89 (1984).

	
\end{thebibliography}
\end{document}